\documentclass[aps,jmp,showpacs,showkeys]{revtex4-1}
\usepackage{graphicx}
\usepackage{dcolumn}
\usepackage{bm}
\usepackage{amssymb,amsmath,latexsym,amsfonts}
\usepackage{color}
\usepackage[dvipsnames]{xcolor}
\usepackage{soul}

\newcommand{\beq}{\begin{eqnarray}}
\newcommand{\eeq}{\end{eqnarray}}

\begin{document}

\title{Superfluids in Polymer Quantum Mechanics}

\author{Jasel Berra--Montiel$^{1,2}$, El\'ias  Castellanos$^{3}$, Alberto Molgado$^{1,2}$ and 
Jonathan Trinidad--Garc\'ia$^1$}
\affiliation{$^1$Facultad de Ciencias \\
Universidad Aut\'onoma de San Luis 
Potos\'i \\  
Campus Pedregal, 
Av.~Parque Chapultepec 1610,
Col.~Privadas del Pedregal,
San Luis Potos\'i­, SLP, 78217, M\'exico}
\affiliation{$^2$Dual CP Institute of High Energy Physics, 
M\'exico\\} 
\affiliation{$^3$Mesoamerican Centre for 
Theoretical Physics\\  Universidad Aut\'onoma de Chiapas\\
Ciudad Universitaria, Carretera Zapata Km. 4, Real del Bosque (Ter\'an), 
29040, Tuxtla Guti\'errez, Chiapas, M\'exico}

\email{jasel.berra@uaslp.mx}
\email{ecastellanos@mctp.mx} 
\email{alberto.molgado@uaslp.mx}

\begin{abstract}
The polymer quantization   stands for  
a non-regular representation of quantum mechanics that results 
non unitarily equivalent to the standard Schr\"odinger representation.  
This quantization is obtained by enforcing the unitary implementation of diffeomorphisms by means of non-regular representation techniques to models with a finite number of degrees of freedom. 
The non-regular nature of 
the polymer representation implies  that only 
unitary Weyl 
operators  are well-defined and, in consequence, in a particular polarization, 
the quantum configuration
space results discrete by the appearance of a minimal-length parameter
(which in the context of loop quantum gravity is usually associated with Planck's 
length).  
In this work, we analyze the corrections obtained on a homogeneous one-dimensional Bose gas within the high densities limit by means of the polymer quantization scheme. Thus, starting from the Bogoliubov formalism,  we analyze the ground expectation value of the polymer momentum operator in terms of semiclassical states, in order 
to obtain an analytic expression for the ground state energy of the N-body system, which allows us to solve the pathological behavior commonly associated with the one-dimensional Bose-Einstein condensation through the introduction of finite size effects characterized by the contribution of the polymer corrections. 
We also discuss
the speed of sound in our polymer version of the Bose gas and the corresponding relative shift induced by the introduction of a minimum length parameter. 
Finally, by considering the idea that the Bose-Einstein condensation phenomenon is closely related to that of superfluidity, we investigate the emergent superfluid behavior in our polymer model by implementing an appropriate Landau's criterion. In this case, we are able to consequently analyze the changes in the critical velocity which defines the limit between the superfluid-condensate regions, thus deducing that the polymer length acts as a kind of pseudo-potential which induces a dissipationless flow associated with the superfluid phase even in the absence of self-interactions.  \\
\end{abstract}


\maketitle

\section{Introduction}

A major aim of the 21st-century physics is to find a unifying theory which provides an accurate description of the considered two cornerstones of modern physics: the behavior of the microscopic world, depicted by the Quantum Theory, on the one hand, and the geometrical interpretation of gravity, covered by General Relativity~\cite{Thiemann}, on the other hand. Even though big efforts have been made over the last few years in order to address this issue, the pursuit of the so-called theory of Quantum Gravity is still an open problem. Research in this direction has been realized from different perspectives such as string theory, supergravity, twistor theory, noncommutative geometry, causal sets, dynamical triangulations, among others~(see review~\cite{Ap1}, and references therein).
However, a framework which has recently attracted some attention from the scientific community  is the one known as Loop Quantum 
Gravity~\cite{Gambini1}, which, when applied to minisuperspace cosmological models leads to a representation known as Loop Quantum 
Cosmology~\cite{Ashtekar}. This approach has lead to notable advances, like the setting of a specific microstate basis for the study of black hole thermodynamics, and also the avoidance of singularities by the existence of a quantum bounce~\cite{Rovelli,Bojowald1}. The quantum representation that naturally emerges within Loop Quantum Cosmology is known as polymer quantization~\cite{Shadow,CTZ}, which stands for  a non-regular implementation of the canonical commutation relations,
obtained through the Weyl unitary operators and for which either 
the position or the momentum are not well-defined as 
operators on a Hilbert space~\cite{Ac3}. Although non-regular representations were overlooked in the past for being considered nonphysical, they have recently acquired interest as they may be useful to treat mechanical systems whose ground states are not depicted by $L^2$ functions~\cite{Cav}. The polymer representation requires the introduction of a minimum-length parameter, called the polymer length, in terms of which the configuration space acquire discrete properties~\cite{Velhinho}. The introduction of a minimal length is also related to the emergence of modified dispersion relations, generalized uncertainty principles (GUP's), and non-commutative geometries~\cite[and references therein]{Gorji,polymer-GUP}, among other implications. Finding experimental evidence on the quantization of space-time is fundamental to validate the theoretical predictions of either Loop Quantum Gravity or Loop Quantum Cosmology.

Within the currently available options to perform experimental measures of  Quantum Gravity predictions, we may enounce the resonator decoherence effect from spontaneous emission, the Talbot-Lau interferometry, the matter-wave interferometry,  and the phenomenon of Bose-Einstein condensation (BEC), among others interesting approaches~\cite{Carney}.  In particular, Bose-Einstein condensation has exceptionally drawn the attention of the scientific community due to its high experimental precision and the possibility of using certain properties of these kinds of systems (condensation temperature~\cite{Castellanos,CastellanosClaus}, the speed of sound~\cite{eli2,ECA}, the free velocity of expansion~\cite{ECA1}, to mention some) as test tools for gravitational physics~\cite{eli,CastellanosCamacho,CastellanosCamacho1,Castellanos,r1,r2,elca,eli2,CastellanosClaus,ECA,ECA1,ECe}. Within the analysis of Bose-Einstein condensates, the lower dimensional models for them result very relevant in the context of condensed matter physics since the nature of their collective excitations and their transition-phase properties are directly dependent on the dimensionality of the system~\cite{Gorlitz}. In particular, an interesting feature related to low-dimensional Bose systems is the impossibility to reach the condensation at finite temperature in the thermodynamic limit~\cite{yuka2}. 

 In order to reach the condensation state in low dimensional Bose gases, it is necessary to take account of finite size effects of the quantum system and, particularly, of its ground state energy.   
Even though a one-dimensional Bose-Einstein condensate can never be achieved in practice, it is possible to control the experimental environment by setting anisotropic traps to obtain a quasi-one-dimensional BEC~\cite{ex1, ex2}.  
Further, in the regime of very low temperatures, it appears another interesting physical phenomenon, known as superfluidity, that is a quantum manifestation in which the ultra-cold matter behaves like a fluid which is able to flow without dissipation of energy, presenting zero viscosity and zero entropy. The common background in which both the superfluid behavior and the Bose-Einstein condensation take place suggests a close relation between these phenomena~\cite{London}, and thus the idea of a simultaneous superfluid-condensate system appears significant and claims for a detailed analysis from the theoretical and the experimental points of view~\cite{Ueda}.  This last issue has been addressed theoretically, for example, through the application of the Landau's criterion to elementary excitations of a Bose gas in order to corroborate if its critical velocity supports the existence of a superfluidity region~\cite{Quintana}.

Bearing this in mind, in this paper we aim to explore the properties of the ground state and the corresponding excitations associated with a homogeneous one-dimensional Bose-Einstein condensate within the polymer quantization program.  To that end, we will start by considering that our polymer system lies in the limit where the Bogoliubov's formalism is assumed to be valid~\cite{CastellanosReg,PQBR}.  In order to obtain well-defined properties for the ground state, we will explore the finite size corrections derived from the calculation of the elementary excitations of the system through semiclassical states considered over regular lattices with a length parameter corresponding to the polymer length. We will determine an analytic expression for the ground state energy of the polymerized BEC, and also we will show the way in which we may recover the standard measures of energy, pressure, and speed of sound, plus polymer finite-size corrections. Further, by exploiting Landau's criterion on excitations of a Bose gas, we will demonstrate that the resultant polymer condensate lies in the superfluid region and also that the existence of a minimum-length parameter implies a shift which reduces the value of the critical velocity. Finally, we will discuss the remarkable issue that the polymer length acts as an interactive pseudo-potential that benefits the existence of a superfluid behavior even in the absence of self-interactions.

The rest of the paper is organized as follows: in Section~\ref{sec:polymer} we present a brief introduction to polymer quantization, putting special 
emphasis on the possible values that the momentum is able to acquire within this quantization framework.  
In Section~\ref{sec:Bogol} we implement the polymer $N$-body 
one-dimensional Hamiltonian for non-relativistic  Bose particles and 
explicitly construct its associated ground energy.
In Section~\ref{sec:sound} we calculate the speed of sound
introduced by considering polymer quantum effects, and fix a  
numerical value for the shift in the speed of sound caused by the polymer 
length.  In Section~\ref{sec:Landau} we analyze the superfluidity phenomenon from the polymer representation perspective.  In particular, 
we discuss in detail the way in which the polymer parameter 
may contribute to the appearance of a superfluid region. 
Finally, in Section~\ref{sec:conclu} we add some concluding remarks.

\section{Polymer Quantization}
\label{sec:polymer}

For the sake of simplicity,   we will only consider quantum systems with one degree of freedom, however, generalization to systems with $n$-degrees of freedom may be straightforward. 
The canonical commutation relations (CCR), also known as Heisenberg commutation relations~\cite{Kadison}, 
\begin{equation}
\label{Heisenberg}
[\hat{x},\hat{p}]:=i\hbar\widehat{\{x,p\}}= i\hbar\hat{I}\,,
\end{equation}
are commonly taken as one of the cornerstones for the standard representation of quantum theory. Nevertheless, it is easy to show that, in order that the CCR~(\ref{Heisenberg}) are satisfied, the quantum operators for position and momentum, $\hat{x}$ and $\hat{p}$, respectively, cannot be both bounded~\cite{Wintner,Wielandt}. In consequence, quantum observables related to such operators cannot be self-adjoint elements defined on the whole Hilbert space. A method to avoid this technical problem was originally introduced by Hermann Weyl~\cite{Weyl}, who noticed that we can map the quantum operators $\hat{x}$ and $\hat{p}$ into one-parametric families of unitary operators which are thus considered as the fundamental operators of the quantum theory~\cite{Stone1}.  
Weyl operators are given by the exponentiated versions of $\hat{x}$ and $\hat{p}$~\cite{PolyPropagators}, namely,
\begin{equation}\label{basic}
\hat{U}(\mu)=e^{i\mu \hat{x}}\,,\qquad \hat{V}(\lambda)=e^{\frac{i}{\hbar}\lambda \hat{p}}\,,
\end{equation}
where $\mu$ and $\lambda$ stand for arbitrary real parameters.  Weyl 
operators~(\ref{basic}) satisfy a product rule which may be obtained through the Baker-Campbell-Hausdorff theorem~\cite{Hall2},
\begin{equation}\label{BCH}
\hat{U}(\mu)\hat{V}(\lambda)=e^{-i\mu\lambda}\hat{V}(\lambda)\hat{U}(\mu)\,.
\end{equation}
Also, they are endowed with an involution $*$-operator,
\begin{equation}
[\hat{U}(\mu)]^*=\hat{U}(-\mu)\,,\qquad [\hat{V}(\lambda)]^*=\hat{V}(-\lambda)\,,
\end{equation}
and fulfill the relations
\begin{eqnarray}\label{Weyl}
&\hat{U}(\mu_{1})\hat{U}(\mu_{2})=\hat{U}(\mu_{1}+\mu_{2})\,,\qquad \hat{V}(\lambda_1)\hat{V}(\lambda_2)=\hat{V}(\lambda_1+\lambda_2)\,.
\end{eqnarray}
Finite linear combinations of the Weyl operators in ($\ref{basic}$) generate a $C^*$--algebra~\cite{Segal, Sakai, Doran}, the algebra of canonical commutation relations (CCR--algebra),  denoted as $\mathcal{A}_w$,
\begin{equation}\label{Walgebra}
\sum_{i}\left(\alpha_{i}\hat{U}(\mu)+\beta_i\hat{V}(\lambda) \right)\in\mathcal{A}_w,\qquad\alpha_i,\beta_i\in\mathbb{C}\,.
\end{equation}
From this perspective,  the quantization procedure is conceived as the assignment of a unitary representation of $\mathcal{A}_w$ in a Hilbert space $\mathcal{H}$. 
Different representations may be obtained from $\mathcal{A}_w$, but 
some of them may be identified by the Stone-von Neumann uniqueness theorem which states that any irreducible representations of the Weyl relations in which the operators are unitary and weakly-continuous result unitarily equivalent to each other and, particularly, to the 
Schr\"odinger representation~\cite{von-Neumann1, Stone2,von-Neumann}.  
Weyl operators hence allow us to recover the infinitesimal generators of transformations if the Stone-von Neumann theorem is satisfied. However, we may relax the weakly-continuous 
condition of this theorem in order to obtain representations that are 
non unitarily equivalent to the Schr\"odinger one.  In this case, the quantum operators $\hat{x}$ and $\hat{p}$ are not recovered as self-adjoint operators in the Hilbert space of the quantum system, and thus their existence is only justified at a heuristic level in order to define the CCR relations.  This situation is commonly referred to as non-regular representations of the CCR-algebra. These representations, which in the past were considered as merely nonphysical pathologies,
have become of interest as they may be completely adapted to analyze 
quantum systems whose ground state does not belong to $L^2$ functions~\cite{Cav}.
In what follows, we will focus our attention in a  particular non-singular representation known as the 
polymer representation of Quantum Mechanics.

One of the most relevant discrepancies between the Schr\"odinger and the polymer representation lies in the choice of a non-separable Hilbert space $\mathcal{H}_{poly}$ in the latter case~\cite{Shadow,CTZ,Fredenhagen}. This Hilbert space is endowed with an uncountable orthonormal basis given by abstract kets $\{|\alpha\rangle\}$, labeled by real numbers such that the inner product satisfies $\langle\alpha|\beta\rangle=\delta_{\alpha,\beta}$, with $\alpha,\beta\in\mathbb{R}$, and $\delta_{\alpha,\beta}$ stands for Kronecker's delta. As a consequence of the non-regularity of the polymer representation~\cite{Ac3, Cav}, the Weyl operators $\hat{U}(\mu)$ and $\hat{V}(\lambda)$, can not be both weakly continuous,  thus excluding the possibility to obtain their respective Hermitian generators, namely the position and momentum operators, as well defined Hermitian operators on 
$\mathcal{H}_{poly}$~\cite{Shadow,CTZ,Fredenhagen,Ac3,Cav}.
Hereinafter, we will  choose to work for convenience in the so-called
$x$-polarization, that is, when the operators $\hat{x}$ and $\hat{V}(\lambda)$ are taken as the fundamental operators for the theory, whereas the momentum operator is not well-defined on $\mathcal{H}_{poly}$~\cite{Morales}. On the one side, the action of these basic 
operators over elements of $\mathcal{H}_{poly}$ in the momentum basis $|p\rangle$ is explicitly given by~\cite{Morales2}
\begin{equation}\label{actions}
\hat{x}|p\rangle=\dfrac{\hbar}{i}\partial_{p}|p\rangle\,,\qquad\hat{V}_{\lambda}|p\rangle=e^{\frac{i}{\hbar}\lambda p}|p\rangle\,,
\end{equation}
where we have defined $\hat{V}_{\lambda}:=\hat{V}(\lambda)$. This means, that in the momentum representation the orthonormal basis is given by plane waves $\langle p|\alpha\rangle=e^{i\alpha p/\hbar}$. The set of linear combination of these plane waves is known as the space of almost periodic functions. Thus, in this basis, we may identify $\mathcal{H}_{poly}$ with the completion of the space of almost periodic functions with respect to the polymer inner product. This space is called the Bohr compactification of the real line  $\bar{\mathbb{R}}_{B}$, and as a locally compact group it is equipped with a translation invariant Haar measure $d\mu_B$. Then, to be more precise, we identify the polymer Hilbert space 
$\mathcal{H}_{poly,p}=L^2(\bar{\mathbb{R}}_{B},d\mu_B)$, which resembles the main features of the Loop Quantum Gravity constructions~\cite{Velhinho, Shadow, Corduneanu}.  On the other side,  the action of the operators $\hat{x}$ and $\hat{V}_{\lambda}$ in the position basis $|x\rangle$ is given by
\begin{equation}
\hat{x}|x\rangle=x|x\rangle\,,\hspace{1em}\hat{V}_{\lambda}|x\rangle=|x-\lambda\rangle\,,
\end{equation}
from this expression we notice that the wave functions of the polymer Hilbert space are tied down to points belonging to a regular lattice, $\{x_j|x_j=x_0+j\lambda,j\in\mathbb{Z}, x_0\in[0,\lambda)\}$ characterized by the parameter $\lambda$, indicating that 
the eigenvalues of the position operator $\hat{x}$ are actually discrete over $\mathcal{H}_{poly,x}$. For a fixed point $x_0$, the elements of the Hilbert space, supported on this lattice, belong to a separable Hilbert subspace of the complete polymer Hilbert space $\mathcal{H}_{poly,x}$, however, in order to determine a consistent physical interpretation of the dynamics, we must restrict the action of Hamiltonian operators to a fixed lattice~\cite{Morales}, as we will see below.

Further, considering that the kinetic energy depends on the expected value of the operator $\hat{p}^2$, which is not a well-defined operator on $\mathcal{H}_{poly}$, and inspired by methods developed in lattice gauge theories, one approximates $\hat{p}^{2}$ in term of the operators $\hat{V}_{\lambda}$ and its inverse 
$\hat{V}_{\lambda}^{\dagger}:=\hat{V}_{-\lambda}$, thus obtaining 
\begin{equation}\label{mom}
\hat{p}^2_{\lambda}=\dfrac{\hbar^2}{\lambda^2}\left[2-\hat{V}_{\lambda}-\hat{V}_{\lambda}^{\dagger} \right]\,.
\end{equation}
By making use of ($\ref{actions}$), we can obtain the eigenvalue for $\hat{p}^2_{\lambda}$ in the momentum basis,
\begin{equation}
\hat{p}^2_{\xi}|p\rangle=\dfrac{\hbar^2}{\xi^2}\sin^2\left(\dfrac{\xi p}{\hbar} \right)|p\rangle\,,
\end{equation}
where we have taken  $\lambda=2\xi$. We will call $\xi$ the polymer length of the system. And thus, from this last eigenvalue equation, we may identify the polymer momentum
\begin{equation}\label{sine}
p_{\xi}:=\dfrac{\hbar}{\xi}\sin\left(\dfrac{\xi p}{\hbar} \right)\,.
\end{equation}

To complete our brief review on the polymer quantization formalism, we need to obtain the expectation value for the previously defined operator. For this purpose, let us work with vacuum states defined over infinite regular lattices $\{x_j|x_j=j\xi, x\in\mathbb{R}, j\in\mathbb{Z}\}$ as prescribed in~\cite{Shadow},
\begin{equation}\label{coherent}
|\psi_0\rangle:=\dfrac{1}{\pi^{1/4}}\sqrt{\dfrac{L}{\xi}}\sum_{j=-\infty}^{\infty}e^{x^2_j/2L^2}|x_j\rangle\,.
\end{equation}
Here, $L \gg \xi$ denotes a length scale parameter introduced in order to equally distribute the uncertainty between the observables $\hat{x}$ and $\hat{p}$ as $\Delta \hat{x}=L/ \sqrt{2}$ and $\Delta \hat{p}=h/\sqrt{2} L$. In other words, we endow the system with a minimum canonical momentum $p=\hbar/\sqrt{2}L$.
Further, by means of the Poisson summation formula~\cite{Benedetto},
\begin{equation}\label{Poisson}
\sum_{n=-\infty}^{\infty}f(x+n)=\sum_{n=-\infty}^{\infty}e^{2\pi i nx}\int_{-\infty}^{\infty}f(y)e^{-2\pi iny}dy\,,
\end{equation}
we are able to determine the approximate expectation values for polymer  
quantum momentum operators.  Hence, 
the expectation values for $\hat{V}_{\xi}$ and $\hat{V}_{\xi}^{\dagger}$ which saturate the uncertainty principle is given by the expression
\begin{equation}
\langle \psi_0|\hat{V}_{\xi}|\psi_0\rangle=\langle \psi_0|\hat{V}_{\xi}^{\dagger}|\psi_0\rangle=e^{-\frac{2\xi^2}{L^2}}\sum_{n=-\infty}^{\infty}e^{\frac{\xi^2}{L^2}[1-2i\pi n\frac{L^2}{\xi^2}]^2}\,.
\end{equation}
We may also note that by taking the condition $L \gg\xi$, these last expressions may be approximated as
\begin{equation}\label{holo}
\langle \psi_0|\hat{V}_{\xi}|\psi_0\rangle=\langle \psi_0|\hat{V}_{\xi}^{\dagger}|\psi_0\rangle=e^{-\frac{\xi^2}{L^2}}\,.
\end{equation}
In terms of ($\ref{holo}$), we may determine the minimum allowed expectation value for $\hat{p}_{\xi}^2$,  namely,
\begin{equation}\label{exp}
\langle \psi_0| \hat{p}^2_{\xi}|\psi_0 \rangle:=\dfrac{\hbar^2}{2\xi^2}\left[1-e^{-\xi^2/L^2} \right]\,.
\end{equation} 
By a simply rearrangement of the terms appearing in ($\ref{exp}$), we are in position to identify that
the minimum expected value allowed for the polymer momentum is given by
\begin{equation}\label{minmom}
p_{\xi_0}:=\dfrac{h}{\xi}e^{-\frac{\xi^2}{4L^2}}\sinh^{1/2}\left({\dfrac{\xi^2}{2L^2}}\right)  \,.
\end{equation}
Taking into account  equations (\ref{sine}) and (\ref{minmom}), it is possible to establish the inequality $p_{\xi_0}\leq p_{\xi}$, from which we are able to see that the minimum canonical momentum associated with a polymer system is given by the expression
\begin{equation}\label{minmomcan}
p_0:=\dfrac{\hbar}{\xi}\arcsin\left[e^{-\frac{\xi^2}{4L^2}}\sinh^{1/2}\left(\dfrac{\xi^2}{2L^2} \right) \right]\,.
\end{equation}
Here,  $p_{0}$ may be interpreted as a semiclassical value of the momentum  at low energies corrected by high energy effects in terms of the polymer scale~\cite{Majumder}. 
In the next section we will analyze the way in 
which the polymer formalism may be implemented
in order to find the expression for the ground 
energy that characterizes the polymer $N$-body
Hamiltonian for a Bose gas.

\section{Polymer quantum mechanics and  Bogoliubov's formalism}
\label{sec:Bogol}

Let us start with the $N$-body one-dimensional regularized Hamiltonian for non-relativistic Bose particles interacting with a momentum independent potential,
\begin{equation}\label{HAM}
\hat{H}=\sum_{p=0} \dfrac{p^2}{2m}\hat{a}^{\dagger}_{p}\hat{a}_{p}+\dfrac{U^{\mathrm{ps}}_x}{2L}\sum_{p=0}\sum_{q=0}\sum_{r=0}\hat{a}^{\dagger}_{q}\hat{a}^{\dagger}_{r}\hat{a}_{q+p}\hat{a}_{r-p}\,.
\end{equation}
Here, $U^{\mathrm{ps}}_x$ stands for the one-dimensional delta-function pseudo-potential, introduced in the standard Bogoliubov theory in order to avoid the divergent $1/x$ behavior from the
scattered wave~\cite{Huang,Olshanii}.  For an arbitrary wave function $\psi$, the pesudo-potential $U^{\mathrm{ps}}_x$ is explicitly given by the operator
\beq
\label{eq:Upseudo}
[U^{\mathrm{ps}}_x(\psi)](x)=U_{0_{1D}}\delta(x)\frac{d}{dx} \left(x \psi(x) \right) \,,
\eeq 
where the term $U_{0_{1D}}$ in~(\ref{eq:Upseudo}) is the potential strength parameter which characterizes the interactions within the system.  We will further describe this term in the following sections.  Besides, $\delta(x)$ stands for Dirac's delta while 
$\frac{d}{dx} \left(x \cdot \right)$ stands for the regularization operator.
Further, $L$ stands for the characteristic length of the one-dimensional condensate, and actually corresponds to the length scale introduced in~(\ref{coherent}) to equally distribute the uncertainty between position and momentum. Finally, the operators $\hat{a}_p$ and $\hat{a}_p^{\dagger}$ in~(\ref{HAM}) denote the standard creation and annihilation operators,
respectively, which satisfy the canonical commutation relations for bosons,
\begin{equation}
[\hat{a}_{p},\hat{a}^{\dagger}_{p'}]=\delta_{p,p'}\,,\qquad[\hat{a}_{p},\hat{a}_{p'}]=0\,,\qquad[\hat{a}^{\dagger}_{p},\hat{a}^{\dagger}_{p'}]=0\,.
\end{equation}
Even though the algebra followed by these operators may result modified within the polymer quantization formalism, we want to introduce an approximation to order in $\xi^{2}$ of the polymer scale, thus allowing us to keep the standard commutation relation for bosons for the operators $\hat{a}$ and $\hat{a}^{\dagger}$~\cite{Husain}.
In order to obtain the ground state energy of the Bose-Einstein condensate with corrections due to the existence of a minimum allowed length scale, we proceed 
by modifying the kinetic energy term in the Hamiltonian ($\ref{HAM}$) with the formal replacement $p\rightarrow p_{\xi}$,
\begin{equation}\label{polyhat}
\hat{H}_{\xi}=\sum_{p=0} \dfrac{\hbar^2}{2m\xi^2}\sin^2\left(\dfrac{\xi p}{\hbar}\right)\hat{a}^{\dagger}_{p}\hat{a}_{p}+\dfrac{U^{\mathrm{ps}}_x}{2L}\sum_{p=0}\sum_{q=0}\sum_{r=0}\hat{a}^{\dagger}_{q}\hat{a}^{\dagger}_{r}\hat{a}_{q+p}\hat{a}_{r-p}\,,
\end{equation}
where we have introduced the polymer momentum~(\ref{sine}).
Henceforth, we will consider the low-energy behavior of the system, that is, 
we may assume that the number of particles in the ground state, $N_{0}$, will be macroscopically occupied, and thus the number of particles in an excited state, $N_{e}$, is small compared with the number of particles in $N_0$. As a consequence, below the condensation temperature we have that $N_0 \approx N$ and  $\sum_{p\neq 0}N_{e}\ll N$, being N the total number of particles.  
%
Bearing in mind this approximation we get $\langle\hat{a}_0^{\dagger}\hat{a}_0\hat{a}_0^{\dagger}\hat{a}_0\rangle\approx N^2$ and $\langle \hat{a}^{\dagger}_0\hat{a}^{\dagger}_0\hat{a}_0\hat{a}_0\rangle\approx N(N-1)\approx N^2$. The weak fluctuations of $\hat{a}_0$  make possible that, to order $N^2$, the field operators $\hat{a}^{\dagger}_0$ and $\hat{a}_0$ commute and behave like c-numbers. This approach is known in the literature as Bogoliubov's approximation \cite{pathria,Ueda}. By extracting the terms in the ground state from the summations in (\ref{polyhat}) up to order $N^2$, the Hamiltonian acquires the form 
\begin{equation}
\label{H}
\begin{split}
\hat{H}_{\xi}=&
\dfrac{\hbar^2}{2m\xi^2}\sin^2\left(\dfrac{\xi p_0}{\hbar}\right)\hat{a}^{\dagger}_{0}\hat{a}_{0}+\sum_{p\neq 0} \dfrac{\hbar^2}{2m\xi^2}\sin^2\left(\dfrac{\xi p}{\hbar}\right)\hat{a}^{\dagger}_{p}\hat{a}_{p}  
+ \dfrac{U^{\mathrm{ps}}_x}{2L}\hat{a}^{\dagger}_{0}\hat{a}^{\dagger}_{0}\hat{a}_0\hat{a}_0\\
&+
\dfrac{U^{\mathrm{ps}}_x}{2L}\sum_{p\neq 0}\left(4\hat{a}^{\dagger}_{0}\hat{a}_0\hat{a}^{\dagger}_{p}\hat{a}_p+\hat{a}_{0}\hat{a}_0\hat{a}^{\dagger}_p\hat{a}^{\dagger}_{-p}+\hat{a}^{\dagger}_{0}\hat{a}^{\dagger}_0\hat{a}_{p}\hat{a}_{-p}\right)\,,
\end{split}
\end{equation}
where $p_0$ in the first term of the right hand side is defined in (\ref{minmomcan}). 
It is important to remark here that the Bose-Einstein condensation phenomenon in one dimension shows a pathological behavior in the thermodynamic limit, namely, if the number of particles is taken to be very large and the length of the system is considered to grow in proportion, keeping the ratio $N/L$ constant, then it is not possible to attain the condensation of the N-body system~\cite{yuka2}. Thus, in order to get a well defined energy, it is necessary to take into account finite size effects of the system which, within our approach, correspond to the value $p_{0} \neq 0$ in the initial value on the summations in (\ref{H}). In other words,  the minimal momentum has to be taken different from zero and must be modified accordingly due to the polymer representation in order to get a well defined energy.  Notice also that we have contributions to the ground state energy given by the first term in the Hamiltonian (\ref{H}). If we set $\xi=0$, we recover the result obtained in \cite{CastellanosReg}, where $p_{0}=\hbar/L$. 
This is an important issue within our formulation since, on the contrary, if we consider a vanishing value for $p_0$, the energy of the system would be divergent and the condensation is never reached at finite temperature.  
Taking into account the normalization condition $N=\hat{a}^{\dagger}_{0}\hat{a}_0+\sum_{p\neq 0}\hat{a}^{\dagger}_{p}\hat{a}_p$, the previous Hamiltonian is simplified to 
\begin{equation}\label{size}
\begin{split}
\hat{H}_{\xi}=&\dfrac{U^{\mathrm{ps}}_x N^2}{2L}+\dfrac{N\hbar^2}{2m\xi^2}e^{-\frac{\xi^2}{2L^2}}\sinh\left(\dfrac{\xi^2}{2L^2}\right)+
\sum_{p\neq 0}\left[\dfrac{\hbar^2}{2m\xi^2}\sin^2\left(\dfrac{\xi p}{\hbar}\right)+\dfrac{U^{\mathrm{ps}}_x N}{L}  \right] \hat{a}^{\dagger}_{p}\hat{a}_{p}\\
&+\dfrac{U^{\mathrm{ps}}_xN}{2L}\sum_{p\neq 0}\left[ \hat{a}^{\dagger}_{p}\hat{a}^{\dagger}_{-p}+\hat{a}_{p}\hat{a}_{-p}\right]\,.
\end{split}
\end{equation}

In order to diagonalize the resultant Hamiltonian, we perform a symplectic linear transformation of the phase space through the application of the so-called Bogoliubov's transformations~\cite{Ueda, pathria},
\begin{equation}\label{Bogoliubov}
\hat{a}_{p}=\dfrac{\hat{b}_{p}-\eta_p\hat{b}^{\dagger}_{-p}}{\sqrt{1-\eta_p^2}}\,,\qquad\hat{a}^{\dagger}_{p}=\dfrac{\hat{b}^{\dagger}_{p}-\eta_p\hat{b}_{-p}}{\sqrt{1-\eta_p^2}}\,,
\end{equation}
where the term $\eta_p$ stands for the Bogoliubov's coefficient, explicitly given by
\begin{equation}
\eta_p=1+\dfrac{L\epsilon_{p_{\xi}}}{U_{0_{1D}}N}-\dfrac{L\epsilon_{p_{\xi}}}{U_{0_{1D}}N}\sqrt{\dfrac{L\epsilon_{p_{\xi}}}{U_{0_{1D}}N}+2}\,.
\end{equation}

Here the term $\epsilon_{p_{\xi}}:=N\hbar^2e^{-\frac{\xi^2}{2L^2}}\sinh\left(\dfrac{\xi^2}{2L^2}\right)/{2m\xi^2}$  
labels the modified energy spectrum for a single particle, and may be written as $\epsilon_{p_{\xi}}=Np_{\xi}^2/2m$.  The operators $\hat{b}_{p}$ and $\hat{b}_{p}^{\dagger}$ stand for the creation and annihilation operators of quasiparticles 
representing elementary excitations of the system and also they obey  the canonical commutation relations for bosons~\cite{pathria}. 

Inserting (\ref{Bogoliubov}) into (\ref{size}) and simplifying, the 
diagonalized Hamiltonian acquires the final form
\begin{equation}\label{diag}
\hat{H}_{\xi}=\dfrac{U^{\mathrm{ps}}_xN^2}{2L}+\epsilon_{p_{\xi}}+\sqrt{\epsilon_{p_{\xi}}\left(\epsilon_{p_{\xi}}+\dfrac{2U^{\mathrm{ps}}_xN}{L}\right)}\hat{b}^{\dagger}_p\hat{b}_p+\sum_{p\neq 0}\left\lbrace -\dfrac{1}{2}\left[\dfrac{U^{\mathrm{ps}}_xN}{L} +\epsilon_{p_{\xi}}-\sqrt{\epsilon_{p_{\xi}}\left(\epsilon_{p_{\xi}}+\dfrac{2U^{\mathrm{ps}}_xN}{L}\right)}\right]\right\rbrace\,,
\end{equation}
where we are able to identify the elementary excitations
\begin{equation}
\label{excitations}
E_{p_{\xi}}=\sqrt{\epsilon_{p_{\xi}}\left(\epsilon_{p_{\xi}}+\dfrac{2U^{\mathrm{ps}}_xN}{L}\right)}\hat{b}^{\dagger}_p\hat{b}_p \,,
\end{equation}
and, in consequence, we have that the energy of the ground state of the system is given by
\begin{equation}
\label{eq:ground-state1}
E_{0_{\xi}}=\dfrac{U^{\mathrm{ps}}_xN^2}{2L}+\epsilon_{p_{\xi}}+\sum_{p\neq 0}\left\lbrace -\dfrac{1}{2}\left[\dfrac{U^{\mathrm{ps}}_xN}{L} +\epsilon_{p_{\xi}}-\sqrt{\epsilon_{p_{\xi}}\left(\epsilon_{p_{\xi}}+\dfrac{2U^{\mathrm{ps}}_xN}{L}\right)}\right]\right\rbrace  \,.
\end{equation}
Let us analyze the $p$-summation in the last expression. Within the polymer formalism, the canonical momentum $p$ attains a maximum value $p=\pi\hbar/2\xi$, so it is not possible to study the behavior of the summation in the limit $p\rightarrow \infty$. Nevertheless, the polymer momentum may tend to infinity if the polymer length is taken to be very small. If we develop an expansion for $p_{\xi}\gg 0$, we will notice that the summation diverges as $(U_{x}^{ps}N/2L)^2/\epsilon_{p_{\xi}}$, a term that goes as $1/p_{\xi}^2$. The Fourier transform of this term brings a $1/x$ factor in the configuration space, so the pseudo-potential will act as $U^{\mathrm{ps}}_x(1/x)=0$, thus removing the divergence. Back into the momentum space, the action of the pseudo-potential may be interpreted as a simple subtraction of the divergent term. For all the other terms which involve the pseudo-potential in $E_{0_\xi}$, we may recall that the Fourier transform of a constant term is just a Dirac delta $\delta(x)$ in the configuration space, so the action of $U_x^{ps}$ over those terms will not provide contributions to the energy corrections and it will be enough to make the replacement $U^{ps}_x\rightarrow U_{0_{1D}}$. 
Hence, the normalized ground state energy of the polymer Bose-Einstein condensate will be depicted by
\begin{equation}\label{pseudocorrection}
\begin{split}
E_{0_{\xi}}=\dfrac{U_{0_{1D}}N^2}{2L}+\epsilon_{p_{\xi}}
+\sum_{p\neq 0}\left\lbrace -\dfrac{1}{2}\left[\dfrac{U_{0_{1D}}N}{L} +\epsilon_{p_{\xi}}-\sqrt{\epsilon_{p_{\xi}}\left(\epsilon_{p_{\xi}}+\dfrac{2U_{0_{1D}}N}{L}\right)}-\dfrac{1}{2\epsilon_{p_{\xi}}}\left(\dfrac{U_{0_{1D}}N}{L} \right)^2\right]\right\rbrace\,,
\end{split}
\end{equation}
where, without loss of generality, the last term is introduced as a renormalization term in order to avoid the divergence appearing  
in the standard $p$-summation.

If we consider a one-dimensional finite box with periodic boundary conditions, the momentum $p$ of the system is quantized as $p=p(j)=2\pi\hbar j/L$, where $j\in\mathbb{Z}$.  Then, a summation over the momenta runs over these integers in such a way that $\sum_{p} f(p)=\sum_{j} f(j)$. Further, by assuming that the energy particle levels are close enough to each other,  we are able to safely  replace the summation in eq.\,(\ref{pseudocorrection}) by an appropriate integration that takes into account the density of states per energy unit \cite{pathria}. Thus we may approximate the summation by the integral, $\sum_{j} f(j)\approx \int dj\,f(j)$, which in terms of the momentum 
reads
\begin{equation}
\sum_p f(p)\mapsto\dfrac{L}{2\pi\hbar}\int f(p)\,dp \,.
\end{equation}

With this in mind, the last line of the Hamiltonian~(\ref{pseudocorrection}) acquires the integral form
\begin{equation}\label{integral}
\begin{split}
I:=-\dfrac{L}{4\pi\hbar}\int_{\mathbb{R}_B} 
\Bigg\lbrace
  & \dfrac{U_{0_{1D}}N}{L}+\dfrac{N\hbar^2}{2m\xi^2}\sin^2\left(\dfrac{\xi p}{\hbar}\right)-\sqrt{\dfrac{N\hbar^2}{2m\xi^2}\sin^2\left(\dfrac{\xi p}{\hbar}\right)\left[\dfrac{N\hbar^2}{2m\xi^2}\sin^2\left(\dfrac{\xi p}{\hbar}\right)+\dfrac{2U_{0_{1D}}N}{L} \right]}   \\
  & -\dfrac{m\xi^2}{N\hbar^2\sin^2\left(\dfrac{\xi p}{\hbar}\right)}\left(\dfrac{U_{0_{1D}}N}{L} \right)^2 
  \Bigg\rbrace\,dp\,,
\end{split}
\end{equation}
where the limits of the $p$-integration lie in the Bohr compactification, with a lower limit corrected by the minimum momentum~($\ref{minmom}$), that is,
\begin{equation}
\label{minp}
\dfrac{\hbar}{\xi}\arcsin\left(e^{-\frac{\xi^2}{4L^2}}\sinh^{1/2}\left(\dfrac{\xi^2}{2L^2} \right) \right)\leq p \leq \dfrac{\pi\hbar}{2\xi}\,.
\end{equation}
In order to simplify the integral~($\ref{integral}$), we will perform a non-canonical transformation from the phase space labeled by the pair $(x,p)$ to the new set of variables $(X,P)$, defined as~\cite{Gorji2}
\begin{equation}
\label{eq:noncanon}
(x,p)\ \ \ \longmapsto \ \ \ \left(X:=x,P:=\dfrac{\hbar}{\xi}\sin\left(\dfrac{\xi p}{\hbar} \right) \right)\,.
\end{equation}
For theories with a generalized uncertainty principle (GUP), the Darboux theorem makes possible to define the Jacobian of the transformation in terms of Poisson brackets~\cite{Gorji3}. In this case, the correspondent Jacobian of the transformation is given by 
\begin{equation}
J=\{X,P\}=\sqrt{1-\left(\dfrac{\xi P}{\hbar} \right)^2}\,,
\end{equation}
in terms of which we find a polymer-deformed density of states
\begin{equation}
\dfrac{L}{2\pi\hbar}\int_{\mathbb{R}_B}dp \ \ \ \longmapsto \ \ \ \dfrac{L}{2\pi\hbar}\int_{\mathbb{R}}\dfrac{dP}{J}\,.
\end{equation}
From the non-canonical transformation~(\ref{eq:noncanon}), it is clear that the domain of the polymer momentum $P$ gets restricted to
\begin{equation}
\dfrac{\hbar}{\xi}e^{-\frac{\xi^2}{2L^2}}\sinh^{1/2}\left(\dfrac{\xi^2}{L^2}\right)\leq P \leq \dfrac{\hbar}{\xi}\,.
\end{equation}
Further, by introducing the change of variable 
\begin{equation}
z:=P\sqrt{\dfrac{L}{2mU_{0_{1D}}N}}\,,
\end{equation}
and judiciously defining the constants
\begin{equation}
\alpha:=e^{-\frac{\xi^2}{4L^2}}\sinh^{1/2}\left(\dfrac{\xi^2}{L^2}\right) \,,\hspace{1cm}\beta:= \dfrac{\xi}{\hbar}\sqrt{\dfrac{2mU_{0_{1D}}N}{L}}\,,
\end{equation}
it is possible to express the ground state energy of the system as 
\begin{equation}\label{grenergy}
E_{0_{\xi}}=\dfrac{U_{0_{1D}}N^2}{2L}+\dfrac{N\hbar^2}{2m\xi^2}e^{-\frac{\xi^2}{2L^2}}\sinh\left(\dfrac{\xi^2}{2L^2}\right)-\dfrac{\sqrt{2m}L}{4\pi\hbar}\left(\dfrac{U_{0_{1D}}N}{L} \right)^{3/2}\int_{\alpha/\beta}^{1/\beta}\frac{dz}{\sqrt{1-\left(\beta z \right)^2}}f(z)\,,
\end{equation}
where the function $f(z)$ is explicitly given by
\begin{equation}
f(z)=1+z^2-z\sqrt{z^2-2}-\frac{1}{2z^2}\,.
\end{equation}
After performing the integral in ($\ref{grenergy}$), we obtain an analytic expression for the ground energy $E_0$ associated with the polymer N-body Hamiltonian:
\begin{equation}\label{analytic}
\begin{split}
E_{0_{\xi}}\,=\,&\dfrac{U_{0_{1D}}N^2}{2L}+\dfrac{N\hbar^2\alpha^2}{2m\xi^2}+\dfrac{U_{0_{1D}}N}{8\pi \xi \beta^2} 
\Bigg\lbrace
   (1+2\beta^2)\Biggl[\arcsin(\alpha)-i\log\left(\dfrac{\sqrt{1-\alpha}-i\sqrt{\alpha^2+2\beta^2}}{\sqrt{1+2\beta^2}} \right)\Biggr]\\
  & - \dfrac{\sqrt{1-\alpha}}{\alpha}\Biggl[\beta^4-\alpha^2-\alpha\sqrt{\alpha^2+2\beta^2} \Biggr]
  \Bigg\rbrace\,.
\end{split}
\end{equation}
Event though this last expression for the polymer ground energy may be completely awkward, in the following section it will serve us to analyze the speed of sound by taking the appropriate limits of the polymer parameter.

\section{Speed of sound}
\label{sec:sound}

In this section we will discuss the speed of 
sound associated with the ground polymer energy
developed above.  Let us start by characterizing the explicit functional form of the potential strength $U_{0_{1D}}$. We assume that the parameter $U_{0_{1D}}$ is given by the expression $U_{0_{1D}}=-2\hbar^{2}/ma_{1D}$, where $a_{1D}$ is the scattering length in one dimension, that is related to the 3 dimensional scattering length $a_{3D}$, as in trapped Bose--Einstein condensates~\cite{PQBR} through the 
expression
\begin{equation}
a_{1D}=-\frac{L^{2}}{2a_{3D}}\Bigl(1-c\frac{a_{3D}}{L}\Bigr)
\end{equation}
where $c$ is a constant that can be determined by experiments. If we assume that $a_{3D}\ll L$ the we are able to write the potential strength as 
\begin{equation}
U_{0_{1D}}\approx 4\hbar^2 a_{3D}/mL^2.
\end{equation}
%
Further, if we develop successive Taylor expansions in the expression for the polymer ground energy~(\ref{analytic}) by considering $\xi \ll 1$ and  $N/L \gg 1$, respectively, we recover the expression \cite{PQBR}
\begin{equation}
\label{eq:E0xi}
\begin{split}
E_{0_{\xi}} \approx & \dfrac{U_{0_{1D}}N^2}{2L}+\dfrac{N\hbar^2}{4mL^2}\left(1-\dfrac{\xi^2}{2L^2}\right)-\dfrac{\sqrt{2m}L}{4\pi\hbar}\left(\dfrac{U_{0_{1D}}N}{L} \right)^{3/2} \left[\dfrac{2\sqrt{2}}{3}-\dfrac{1}{2g}-g \right]\\
& + \dfrac{\sqrt{2m}}{240\pi\hbar L}\left(\dfrac{U_{0_{1D}}N}{L} \right)^{3/2}\left[\dfrac{5}{g}-32 g^2 \right]\xi^2\,,
\end{split}
\end{equation}
where, for simplicity, we have introduced the definition
\begin{equation}
g:=\dfrac{\beta L}{\sqrt{2}\xi} =\dfrac{\sqrt{mU_{0_{1D}}NL}}{\hbar}\,.
\end{equation}

%
%
%

From expression~(\ref{eq:E0xi}), we see that 
in the limit $\xi\to 0$ we are able to recover the usual ground state energy corrected by finite size terms labeled by the constant $g$ \cite{CastellanosReg}.  Further, from  the approximated polymer ground state 
energy~(\ref{eq:E0xi}), we may also determine the corresponding pressure of the condensed gas, 
\begin{equation}
\begin{split}
P_{\xi}:=&-\left( \frac{\partial E_{0_\xi}}{\partial L}\right)\\
=& \dfrac{U_{0_{1D}}N^2}{2L}+\dfrac{N\hbar^2}{2mL^3}\left(1-\dfrac{\xi^2}{L^2} \right)-\dfrac{\sqrt{2m}}{8\pi\hbar}\left(\dfrac{U_{0_{1D}}N}{L} \right)^{3/2} \left[\dfrac{2\sqrt{2}}{3}-\dfrac{1}{g}\right]\\
& + \dfrac{\sqrt{2m}}{80\pi\hbar L^2}\left(\dfrac{U_{0_{1D}}N}{L} \right)^{3/2}\left[\dfrac{5}{g}-16 g^2 \right]\xi^2\,,
\end{split}
\end{equation}
and the squared speed of sound within the body, namely,
\begin{equation}
\begin{split}
c_{\xi}^2:=& -\frac{L^2}{Nm} \left( \frac{\partial E_{0_{\xi}}}{\partial L} \right)\\ =& \dfrac{U_{0_{1D}}N}{mL}+\dfrac{3\hbar^2}{2m^2L^2}\left(1-\dfrac{5\xi^2}{3L^2} \right)-\dfrac{\sqrt{2m}L}{4\pi\hbar m N}\left(\dfrac{U_{0_{1D}}N}{L} \right)^{3/2} \left[\dfrac{1}{\sqrt{2}}-\dfrac{1}{g}\right]\\
& + \dfrac{\sqrt{2m}}{4\pi\hbar m L}\left(\dfrac{U_{0_{1D}}N}{L} \right)^{3/2}\left[\dfrac{1}{g}-2 g^2 \right]\xi^2\,.
\end{split}
\end{equation}

Although, as mentioned before,  from the experimental perspective it is not possible to obtain a one-dimensional condensate, the construction of a quasi-one-dimensional 
condensate is attainable through the implementation of either optical dipole traps or highly anisotropic magnetic traps, which allow to freeze out the oscillations at low temperatures, thus providing an effective one-dimensional 
model~\cite{Haugset}.
In order to analyze the sensitivity of the system to corrections obtained by the polymer scheme of quantization and its possible behavior into the superfluid region, let us consider the conditions for the theoretical model of a sodium gas
for which the following conditions follows~\cite{Quintana}: $N\propto10^6$ particles, $ a_{3D}\propto10^{-9}$ m, $m\propto10^{-26}$ kg and $L\propto10^{-4}$ m.  Such conditions allow us to make an estimation of the relative shift produced by $\xi$,
\begin{equation}
\label{ssv}
\Delta c_{\xi}^2:=\dfrac{c_{\xi}^2-c_{\xi=0}^2}{c_{\xi=0}^2}\approx - \xi^2\,6.03 \times 10^6,
\end{equation}
whose form indicates a reduction of order $10^6$ in the standard value of the speed of sound. Of course, in the limit $\xi\to 0$, this correction vanishes. If we take into account the bound $\xi^{2} \propto 10^{-16}$ ~\cite{eli}, the relative shift obtained has an order of $10^{-10}$ m$^2$/s$^2$, that is, at least seven orders of magnitude below the standard value under experimental conditions. Nevertheless, keeping the ratio $N/L$ constant and tuning the scattering length to higher values would lead to a remarkable improvement on this calculation. If we set, for example, $a_{3D}\propto10^{-3}$ m, the shift increases to $-5.65\times10^{-4}$ m$^2$/s$^2$, a quantity that results  more reasonable to measure from the experimental point of view. The modification of the scattering length may be realized, for example, by performing a tuning with Feshbach resonances~\cite{Lippe}. 

\section{Modified Landau's criterion}
\label{sec:Landau}

Superfluidity is a microscopical phenomenon in which a fluid under certain critical velocity is able to flow without energy dissipation~\cite{Schmitt, Raman}. Let us consider a condensate sitting at rest in a pipe. If we apply a boost with velocity $v$ in the positive $x$-direction, the imperfections in the pipe will induce back-scattering in the particles of the condensate, dissipating energy. However, from the perspective of the rest frame of the fluid, one may translate this effect as the appearance of excitations with energy $E_p$ and  momentum $p$ moving in the opposite direction. A Galilean transformation and the application of the energy conservation principle allow us to observe that the change in energy between the frame in rest and the moving frame is given by $\Delta E=E_p-v\cdot p$. In order to have an exchange of energy with the pipe, the condition $E_p-v\cdot p <0$ is necessary, and it may only be satisfied if $|v|<E_p/|p|$.
Hence, if the flow velocity $v$ of a condensate with elementary excitations $E_p$ stays under the critical velocity $v_c$, determined from the condition
\begin{equation}\label{Landau}
v_c=\min_p \dfrac{E_p}{|p|}\,,
\end{equation}
the N-body system will be able to flow without any dissipation at all~\cite{Raman, Bobrov}. If the critical velocity is reached, then the superfluid phase will be broken, bringing
into play the production and growth of excitations~\cite{Annet, Legget}, as has been experimentally shown by studying either superfluid helium~\cite{Allum},
ultracold bosons~\cite{Raman} or fermions~\cite{Miller}.
Groundbreaking Bogoliubov's calculation of the properties of a dilute condensate showed that the interplay between the Bose-Einstein condensation and the interatomic interactions produces a linear dispersion of the elementary excitations, which according to Landau's criterion,  exhibits a dissipationless flow when the group velocity of the condensate is smaller than the speed of sound~\cite{Astrakharchik,Ianesell}.
In order to apply Landau's criterion into our condensate, we need to perform slight modifications in order to make it compatible with the polymer formalism,
as developed above. Indeed, exploiting the formal replacement $p\rightarrow p_{\xi}$,  taking into account the dependence of $p_{\xi}$ on $p$ (that is, the minimum of the polymer momentum $p_{\xi}$ is obtained by minimizing the canonical momentum $p$) and, finally, bearing in mind that the polymer formalism has provided us already with a non-zero minimum value for $p_{\xi}$, 
we can propose within our polymer setup a critical velocity, $v_{c_{\xi}}$, which may be obtained  by recasting  Landau's criterion as
\begin{equation}
v_{c_{\xi}}:=\dfrac{E_{0_{\xi}}}{|p_{{\xi}_0}|}\,,
\end{equation}
where we have to consider the energy for the elementary excitations~(\ref{excitations}) in the appropriate limit, that is, the expression 
\begin{equation}
E_{0_{\xi}}=\sqrt{\epsilon_{0_{\xi}}\left(\epsilon_{0_{\xi}}+\dfrac{2U_{0_{1D}}N}{L}\right)}
\end{equation}
that
corresponds with the minimum elementary excitations of the polymer-corrected system.
Keeping this in mind, we determine the value of the critical velocity for our polymer system as
\begin{equation}\label{Critical}
\begin{split}
v_{c_{\xi}}=\sqrt{\dfrac{\hbar^2}{4m^2\xi^2} e^{-\frac{\xi^2}{2L^2}}\sinh\left(\dfrac{\xi^2}{2L^2}\right)+\dfrac{U_{0_{1D}}N}{mL}}\,.
\end{split}
\end{equation}
Performing a Taylor's expansion around $\xi=0$, the last expression acquires the form
\begin{equation}
v_{c_{\xi}}=s\sqrt{1+\gamma^2}-\dfrac{s\gamma^2}{\sqrt{1+\gamma^2}}\left(\dfrac{\xi}{2L} \right)^2\,,
\end{equation}
where we have introduced 
$s:=\sqrt{U_{0_{1D}}N/mL}$ (the usual speed of sound) and also we have defined the constant $\gamma=\hbar/2 \sqrt{2} Lms$. As before, we are able to determine the relative shift in the critical velocity, due to 
corrections induced within the polymer framework,
\begin{equation}
\frac{\Delta\, v_{c_{\xi}}}{{v_{c_{\xi=0}}}}:= \dfrac{v_{c_{\xi}}-v_{c_{\xi=0}}}{v_{c_{\xi=0}}}=-\dfrac{\gamma^2}{1+\gamma^2}\left(\dfrac{\xi}{2L} \right)^2,
\end{equation}
By using the same experimental conditions for the shift in the speed of sound for a sodium gas~(\ref{ssv}), we obtain for the relative shift in the critical velocity
\begin{equation}
\frac{\Delta\, v_{c_{\xi}}}{{v_{c_{\xi=0}}}}=-3.9\times10^4\xi^2\,.
\end{equation}
We can observe a resultant negative shift whose effect is to diminish the superfluid region where the condensate begins.  
Further, 
by going back to the expression for the critical velocity~(\ref{Critical}), and in
 order to study an scenario in which the 
self-interaction parameter $U_{0_{1D}}$ vanishes (that is, the ideal case), we notice that the critical velocity reads as
\begin{equation}
v_{c_{\xi}}=\dfrac{\hbar}{2m\xi}e^{-\frac{\xi^2}{4L^2}}\sinh^{1/2}\left(\dfrac{\xi^2}{2L^2}\right)\,,
\end{equation}
which may be approximated in the limit $\xi\to 0$ as
\begin{equation}
v_{c_{\xi}}=\dfrac{\hbar}{2\sqrt{m}L}\left[1-  \dfrac{\xi^2}{2L^2}\right]\,,
\end{equation}
The relative shift in this case is given by
\begin{equation}
\frac{\Delta\, v_{c_{\xi}}}{{v_{c_{\xi=0}}}}=-\dfrac{\xi^2}{2L^2}\,,
\end{equation}
a result implying that, even in a scenario in which the self-interaction parameter vanishes, the polymer correction induces a shift that solely depends on  the characteristic length of the 
one-dimensional condensate $L$ and on the polymer length $\xi$. 
In other words, we have induced the superfluidity phenomenon as a consequence of the existence of a discrete structure of the background space-time by means of the polymer length $\xi$ even in the absence of a pseudo-potential, in which case, the critical velocity has an explicitly dependence on the minimum length physically allowed while 
the relative shifts in the critical velocity of the superfluid remains negative, therefore the effect of the polymer quantization correction is a contribution which diminishes the critical velocity of the polymer superfluid-condensate.


\vspace{5ex}

\section{Conclusions}
\label{sec:conclu}

Taking into account Bogoliubov's formalism as a guiding principle, we have analyzed some properties associated with the ground state of a homogeneous one-dimensional Bose-Einstein condensate under a non-regular representation of Quantum Mechanics known as the polymer quantization scheme.  This non-regular representation is characterized by a discrete quantum configuration
space associated with a minimal-length parameter, which has as a consequence a modified minimum momentum in the system.  
Outstandingly, this minimum momentum results completely suited in the case under study, as it allows us to circumvent the pathological behavior which arises in the low-dimensional system through finite-size contributions, along with corrections of the order of the polymer length scale. In consequence, we were able to obtain a correction to the standard analytic expression for the ground energy, which leads us to an approximated identification of the speed of sound.   Our results allowed us to obtain a precise estimation of the relative shift on the speed of sound induced by the polymer length scale.  We claim that this estimation may even be experimentally measured by implementing appropriate modifications to the scattering length. Further,  within the polymer quantization formalism, we have modified and implemented Landau's criterion in order to obtain the critical velocity which establishes a limit for the Bose-Einstein condensate to behave as a superfluid. As discussed above, the polymer corrections in the system produce a shift that diminishes the value of this critical velocity, even in the case when the self-interactions of the system vanish, which is quite remarkable.   In consequence, we may interpret the corrections caused by the polymer length as some type of pseudo-interaction caused by the discrete nature of the quantum structure of space-time.  Although the results presented in this work were obtained for a one-dimensional Bose gas, we assert that similar results may be achieved in a more realistic situation, for example in the analysis of an inhomogeneous system (trapped) within the Bogoliubov formalism and also extended to higher spatial dimensions.  This last point deserves further investigation and will be presented elsewhere.

Finally, even though polymer quantization is widely used inside the Loop Quantum Gravity context and, in particular, within its finite-dimensional counterpart in the Cosmological framework, its applicability in other areas of Physics is not completely developed.  We hope that our work also contributes to motivating the study of non-regular representations of Quantum Mechanics in other contexts where they can surely be relevant due to their attributes. 

\begin{acknowledgments}
EC and JTG  acknowledges UNACH/MCTP for financial support.  
JTG, JBM and AM would like to acknowledge financial support from CONACYT-Mexico under projects CB-2014-243433 and CB-2017-283838.
\end{acknowledgments}

\end{document}